
\hsize=15.5 truecm
\vsize=22.5 truecm
\leftskip=1 truecm
\topskip=1 truecm
\splittopskip=1 truecm
\parskip=0 pt plus 1 pt
\baselineskip=18.9 pt

\def\a{\alpha}
\def\b{\beta}
\def\d{\delta}
\def\g{\gamma}

\def\e{\epsilon}

\def\P{\Psi}
\def\o{\omega}

\def\O{\Omega}
\def\n{\nabla}
\def\S{\Sigma}
\def\s{\sigma}

\bigskip
\centerline{\bf MAXWELL FIELDS IN SPACETIMES ADMITTING}
\centerline{\bf  NON-NULL KILLING VECTORS }
\bigskip
\bigskip
\centerline{{Istv\'an R\'acz}}
\medskip
\centerline{MTA KFKI Research Institute for Particle \& Nuclear
Physics}
\centerline{H-1525 Budapest 114, P.O.B. 49, Hungary }
\bigskip
\bigskip\parindent 0pt

 {\bf Abstrac:} We consider source-free electromagnetic fields  in
spacetimes possessing  a non-null  Killing vector  field, $\xi^a$.
We assume further that the electromagnetic field tensor, $F_{ab}$,
is invariant  under the  action of  the isometry  group induced by
$\xi^a$.  It is proved that whenever the two potentials associated
with the  electromagnetic field  are functionally  independent the
entire  content  of  Maxwell's  equations  is  equivalent  to  the
relation  $\n^aT_{ab}=0$.   Since  this  relation  is  implied  by
Einstein's equation  we argue  that it  is enough  to solve merely
Einstein's equation  for these  electrovac spacetimes  because the
relevant equations of motion will be satisfied automatically.   It
is  also  shown  that  for  the  exceptional  case of functionally
related potentials  $\n^aT_{ab}=0$ implies  along with  one of the
relevant  equations  of  motion  that  the  complementary equation
concerning the electromagnetic field is satisfied.

\bigskip
PACS numbers: 04.20.Cv, 04.20.Me, 04.40.+c
\bigskip
\bigskip
\parindent 20pt

 In the  framework of  general relativity  the spacetime geometry,
which serves  as a  stage for  the history  of physical fields, is
influenced  by  the  matter   content  of  the  spacetime.    More
definitely, the geometry of a  spacetime is related to the  matter
distribution  by  Einstein's  equation $$R_{ab}-{1\over 2}g_{ab}R=
8\pi  T_{ab},\eqno(1)$$  where  $R_{ab}$  denotes the Ricci tensor
associated to $g_{ab}$ and $T_{ab}$ is the energy-momentum  tensor
of matter  fields.  In  fact, when  one is  looking for spacetimes
which are solutions  of Einsten's equation  with selected sort  of
matter fields beside eq.   (1) one has to  consider simultaneously
the relevant Euler-Lagrange  equations governing the  evolution of
matter fields, as well.   In general, however, these  equations of
motion for the matter fields and Einstein's equation give rise  in
a  selected  local  coordinate  system  to a complicated system of
coupled non-linear second order partial differential equations for
the  components  of  the  spacetime  metric  and the tensor fields
representing the  matter fields  in the  spacetime.  On  the other
hand,   Einstein's   equation   --   or   more   precisely,    its
``integrability condition", the twice contracted Bianchi  identity
-- implies the relation $$\n^aT_{ab}=0,\eqno(2)$$ which contains a
great deal of information on the behaviour of matter fields.  More
definitely,  the  above  relation  gives  rise,  in  general,   to
algebraical relationships  between various  terms of  the relevant
equations  of  motion  (see,  for  instance,  Ref.  [1]),  and  so
Einstein's equation implies along with some of the  Euler-Lagrange
equations  that  the  complementary  equations  of  motion for the
matter fields  are satisfied.   In particular,  the Euler-Lagrange
equations are for a perfect  fluid source equivalent to eq.   (2).
Moreover, it can easily be  checked that the whole content  of the
equations of motion is also  equivalent to the above relation  for
an arbitrary real scalar field.   Hence, for these sort of  matter
fields it is enough to  postulate the form of the  energy-momentum
tensor, $T_{ab}$, and solve Einstein's equation alone because then
the relevant equations of motion concerning the evolution of these
matter fields are automatically satisfied.

 The  purpose  of  this  paper  is  to  show  that in a spacetime,
$(M,g_{ab})$, possessing a non-null Killing vector field, $\xi^a$,
and  an  antisymmetric   tensorfield,  $F_{ab}$,  representing   a
source-free electromagnatic  field invariant  under the  action of
the isometry group induced  by $\xi^a$ Maxwell's equations  are
just  consequences  of  the  relation  $\n^aT_{ab}=0$ whenever the
associated two potentials are functionally independent.  Hence, in
particular, for these electrovac  spacetimes it is also  enough to
prescribe   the   form   of   $T_{ab}$   known   for   source-free
electromagnetic  fields  and  solve  then  the relevant Einstein's
equation.  The exceptional case of functionally related potentials
is   also   considered   and   it   is   shown   that the relation
$\n^aT_{ab}=0$ together  with half  of the  relevant equations  of
motion imply that the complementary equation holds.

 Before giving the proof of the above statements we recall some
notions and  results.   In a  spacetime,
$(M,g_{ab})$, the electromagnetic field  can be represented by  an
antisymmetric  tensor  field,  $F_{ab}$,  and the
electric  and  magnetic  fields  measured  by  an  observer   with
4-velocity, $u^a$, are given as [2]
$$E_a=F_{ab}u^b \ \ \ \ {\rm and }\ \ \ \ \ \ B_a=-{1\over 2}
\epsilon_{abcd}\ u^bF^{cd},\eqno(3)$$ where $\e_{abcd}$ denotes
the volume element associated with $g_{ab}$.
The Maxwell's equations for a source-free
electromagnetic field take the following form
$$\nabla^a F_{ab}=0,\eqno(4)$$
$$\nabla_{[a} F_{bc]}=0,\eqno(5)$$
where $\n^a$ denotes the covariant derivative operator determined
by $g_{ab}$. Since we have a preferred non-null vector field,
$\xi^a$, on $M$ it
is advantageous to apply the following decomposition of $F_{ab}$:
$$F_{ab}=-(\xi^e\xi_e)^{-1}\bigl\{2\xi_{[a}\P_{b]}+
\epsilon_{abcd}\xi^c\O^d\bigr\},\eqno(6)$$
where
$$\P_a=F_{ab}\xi^b \ \ \ \ {\rm and }\ \ \ \ \ \O_a=-{1\over 2}
\epsilon_{abcd}\xi^bF^{cd}.\eqno(7)$$
Note that whenever the Killing field, $\xi^a$, is timelike $\P_a$
(resp. $\O_a$) is proportional to the electric (resp. magnetic)
field measured by observers moving along the Killing
trajectories.

To present a more substantial description of the electromagnetic
field it is worth introducing the notion of duality. The dual of a
2-form, $X_{ab}$, is defined as
$$\widetilde X_{ab}={1\over 2}\e_{abcd}X^{cd}.\eqno(8)$$
Then, $\O_a$ can be given in terms of $\xi^a$ and $\widetilde
F_{ab}$
as $$\O_a=-\widetilde F_{ab}\xi^a,\eqno(9)$$ as well as, eqs. (4)
and (5) can be recast into the equivalent form
$$\nabla_{[a} \widetilde F_{bc]}=0,\eqno(10)$$
$$\nabla^a \widetilde F_{ab}=0.\eqno(11)$$
Consequently, the Maxwell's equations are for a source-free
electromagnetic field equivalent to
either of the pairs of eqs. (4) and (11), or (5) and
(10).

Consider, now, the second
pair of these equations, i.e., $$\nabla_{[a} \widetilde
F_{bc]}=0,\eqno(12)$$ $$\nabla_{[a} F_{bc]}=0.\eqno(13)$$
Remember that we have a preferred vector field, $\xi^a$, on the
spacetime and so,
by the application of the projection operator, ${h^b}_a$,
associated with $\xi^a$, i.e., the operator
 ${h^b}_a={\delta^b}_a -(\xi^e\xi_e)^{-1}\xi^b\xi_a$, we may
uniquely decompose the 3-form, $\n_{[a}F_{bc]}$, into 3-forms such
that they have definite ``tangential" or ``perpendicular"
character with regard to their free indices.
Clearly,  $\nabla_{[a} F_{bc]}$ coincides then with the
zero 3-form if and only if its various types of projections
vanish. It is easy to check that there are only two
non-identically vanishing projections of $\nabla_{[a} F_{bc]}$,
namely, $\nabla_{[e} F_{fg]}{h^e}_a {h^f}_b {h^g}_c$ and
$\nabla_{[e} F_{fg]}{h^e}_a {h^f}_b \xi^g$ or,
equivalently, $\nabla_{[a} F_{be]}\xi^e$.
Therefore Maxwell's
equations are equivalent to the following set of equations
$$\nabla_{[e} F_{fg]}{h^e}_a {h^f}_b {h^g}_c=0,\eqno(14)$$
$$\nabla_{[e} \widetilde F_{fg]}{h^e}_a {h^f}_b {h^g}_c=0,\eqno(15)$$
$$\nabla_{[a} F_{be]}\xi^e=0,\eqno(16)$$
$$\nabla_{[a} \widetilde F_{be]} \xi^e=0.\eqno(17)$$

It is well-known that, for an arbitrary 2-form,
$X_{ab}$, the identity
$$2\nabla_{[a}\Phi_{b]}= 3\nabla_{[a}X_{be]}\xi^e-{\cal
L}_\xi X_{ab},\eqno(18)$$ with
$\Phi_a= X_{ab}\xi^b$ is satisfied, where ${\cal L}_\xi X_{ab}$
denotes the Lie derivative of $X_{ab}$ with respect to $\xi^a$.
Applying this identity
for the 2-forms, $F_{ab}$, and ,$\widetilde F_{ab}$, -- which are,
according to our assumption, invariant
under the action of the isometry group associated with the Killing
field, $\xi^a$, -- we get that eqs. (16) and (17) are satisfied if
and only if the 1-forms, $\P_a$, and, $\O_a$, defined by eq.
(7) are closed, i.e.,
whenever there are functions at least locally, $\P$, and, $\O$,
such that $\P_a=\nabla_a\P$ and $\O_a=\nabla_a\O$. Note that, as a
consequence of the definition $\P_a$ and $\O_a$, we have
$${\cal L}_\xi \P={\cal L}_\xi \O=0.\eqno(19)$$

The remaining two equations of motion, i.e., eqs. (14) and (15),
can be shown to be
equivalent to
$$(\nabla^a  F_{ab})\xi^b=0,\eqno(20)$$
$$(\nabla^a \widetilde F_{ab})\xi^b=0,\eqno(21)$$
from which, by the substitution of the decomposition
of $F_{ab}$ in
terms of $\P$, $\O$ and $\xi^a$, we obtain
[3,4] $$\n^a\n_a \P -v^{-1}\bigl\{(\n^a\P)(\n_a
v)+(\n^a\O)\o_a\bigr\}=0,\eqno(22)$$
$$\n^a\n_a \O -v^{-1}\bigl\{(\n^a\O)(\n_a
v)-(\n^a\P)\o_a\bigr\}=0,\eqno(23)$$
where $v=\xi^e\xi_e$ and $\o^a$ denotes the twist vector of
$\xi^a$, i.e., $\o^a=\e^{abcd}\ \xi_b\n_c\xi_d$.

Summarizing the above results we can say that in a spacetime
possessing a non-null Killing vector field, $\xi^a$,  a 2-form,
$F_{ab}$, which is invariant under the action of the isometry
group
induced by the Killing field satisfies Maxwell's equations for a
source-free electromagnetic field if and only if  there exist at
least locally two functions, $\P$, and, $\O$, on $M$ such that
${\cal L}_\xi \P={\cal L}_\xi\O=0$ and
$$F_{ab}=-2v^{-1}\Bigl\{\xi_{[a}
\n_{b]} \P +(\xi_{[a} \n_{b]} \O){\ }\widetilde { }{\ }\Bigr\},
\eqno(24)$$ as well as, eqs. (22) and (23)
are satisfied.

 In  the  remaining  part  of  this  paper  --  using  the   above
formulation  of  electrovac  fields  in  spacetimes  admitting   a
non-null  Killing  vector  field  --  we  are going to examine how
strong restrictions on the relevant equations  of motion can  be
derived from the relation $\n^aT_{ab}=0$.

Consider, now, the energy-momentum tensor of an electrovac
Maxwell field, i.e.,
$$T_{ab}={1\over 4\pi}\Bigl\{F_{ac}{F_b}^c - {1\over
4}g_{ab}F_{ed}F^{ed}\Bigr\}.\eqno(25)$$
The substitution of the decomposition of $F_{ab}$ in terms of
$\P$, $\O$ and $\xi^a$ into eq. (25) yields
$$\eqalign{T_{ab}={v^{-2}\over
4\pi}\Bigl\{&\bigl[(\n^e\P)(\n_e\P)+(\n^e\O)(\n_e\O)\bigr](\xi_a\xi_b-
{1\over 2}v
g_{ab})\cr&+v\bigl[(\n_a\P)(\n_b\P)+(\n_a\O)(\n_b\O)\bigr]+
2\xi_{(a}\e_{b)cgh}(\n^c\P)\xi^g(\n^h\O)\Bigr\}.\cr}\eqno(26)$$
Then, using the fact that $\xi^a$ is a Killing field -- and so,
for instance, $\n_{(a}\xi_{b)}=\xi^a\n_a v=\n^a\xi_a=0$ -- ,
furthermore,
that ${\cal L}_\xi \P={\cal L}_\xi \O=0$, by a straightforward
calculation we obtain
$$\eqalign{\n^aT_{ab}={v^{-1}\over
4\pi}\Bigl\{&\bigl[\n^a\n_a\P-v^{-1}(\n^a\P)(\n_av)\bigr](\n_b\P)\cr+&
\bigl[\n^a\n_a\O-v^{-1}(\n^a\O)(\n_av)\bigr](\n_b\O)\Bigr\}+
{1\over 4\pi}\n^a\bigl[2v^{-2}\xi_{(a}\S_{b)}\bigr]
,\cr}\eqno(27)$$
where $\S^a=\e^{abcd}(\n_b\P)\xi_c(\n_d\O)$ which is for the case
of a timelike Killing field
proportional to the Poynting vector, $P^a={1\over
2}(-v)^{-1/2}\e^{abcd}E_bB_c\xi_d$, of the electromagnetic field
measured by observers moving along the Killing trajectories.

Note that whenever $(\n_{[a}\P)(\n_{b]}\O)=0$ holds in an open
region, i.e., $\n_a\P$ and
$\n_a\O$ are proportional, $\S_a$ vanishes there identically.
Then, either $\P$ or $\O$ is constant, or neither $\n_a\P$
nor
$\n_a\O$ vanishes in the selected open region. In the first
case, eq. (27) simplifies into
$$\n^aT_{ab}={v^{-1}\over
4\pi}\bigl[\n^a\n_a\Phi-v^{-1}(\n^a\Phi)(\n_av)\bigr]
(\n_b\Phi),\eqno(28)$$
where $\Phi$ denotes either of the fields, $\P$, or, $\O$,
according to that we are
considering a pure ``electric" or ``magnetic" field. This
conclusion
completes the proof of our conjecture for this particular case.
For the second case, note that if there is an open region where
$\n_a\P$ and  $\n_a\O$ are non-vanishing the potentials, $\P$ and
$\O$, are functionally related. Then, it
can easily be seen that
$(\n_b\O)\o_a(\n^a\P)-(\n_b\P)\o_a(\n^a\O)
=0$ holds which yields along with eq. (27)
$$\eqalign{\n^aT_{ab}={v^{-1}\over
4\pi}\Bigl\{&\bigl[\n^a\n_a\P-v^{-1}\bigl\{(\n^a\P)(\n_av)
+ \o_a\n^a\O\bigr\} \bigr](\n_b\P)\cr+&
\bigl[\n^a\n_a\O-v^{-1}\bigl\{(\n^a\O)(\n_av)-\o_a\n^a\P
\bigr\}\bigr](\n_b\O)\Bigr\} .\cr}\eqno(29)$$
Hence for this particular case, the relation $\n^aT_{ab}=0$ alone
does not imply that Maxwell's equations are satisfied.
Nevertheless, we may economize the content of eq. (29) by arguing
that the relation $\n^aT_{ab}=0$ together with one of eqs. (22)
and (23) always
implies that the complementary equation of motion is
automatically satisfied.

 For  the  remaining  part  of  this  proof,  we shall assume that
$\n_a\P$ and $\n_a\O$ are  linearly independent.  Note that  in an
open  region  of  $M$  where  $\n_a  \P$ and $\n_b\O$ are linearly
independent the four 4-vectors, $\xi^a,\n^a\P,\n^a\O$ and  $\S^a$,
give     rise     to      a     well-defined     basis      field,
$\{\xi^a,\n^a\P,\n^a\O,\S^a\}$, provided that $\S^a$ is not a null
vector field which may occur when $\xi^a$ is spacelike.  If $\S^a$
happens  to  be  null  it  can  be  shown  that either $\n^a\P$ or
$\n^a\O$ has to be parallel  to $\S^a$.  Denote by $\Phi$  the one
from $\P$ and $\O$ for which $\n^a\Phi$ is proportional to  $\S^a$
and  by  $\Phi^*$  the  other  electromagnetic  potential.   Then,
instead of the above basis field, we may use the pseudo-orthogonal
basis field, $\{\xi^a,\n^a\P,\n^a\O,\s^a\}$, where $\s^a$  denotes
the  unique  null  vector  field  for  which $\s^a\n_a\Phi=-1$ and
$\s^a\xi_a=\s^a\n_a\Phi^*=0$ hold throughout.

Consider, now, the last term of the right hand side of eq. (27).
It can easily be checked
$$\n^a\bigl[2v^{-2}\xi_{(a}\S_{b)}\bigr]=
v^{-2}\Bigl\{\xi_a\n^a\S_b+ \S_a\n^a\xi_b+ \xi_b\n^a\S_a
-2v^{-1}\xi_b\S_a(\n^av)\Bigr\}.\eqno(30)$$
On the other hand, $\n^a\bigl[2v^{-2}\xi_{(a}\S_{b)}\bigr]$
can uniquely be decomposed, with regard to the basis field
$\{\xi^a,\n^a\P,\n^a\O,{\widehat \S}^a\}$
where ${\widehat
\S}^a$ denotes either $\s^a$ or $\S^a$ according to that $\S^a$
is null or not, as
$$\n^a\bigl[2v^{-2}\xi_{(a}\S_{b)}\bigr]=\a \xi_b +\b (\n_b\P)+\g
(\n_b\O)+\d{\widehat \S}_b.\eqno(31)$$
Then, with the aid of eq. (31), using the fact that $\xi^a$ is a
Killing field, and the orthogonality properties
of the vector system $\{\xi^a,\n^a\P,\n^a\O,{\widehat \S}^a\}$,
we find
$$\a=v^{-2}\bigl(\n^a\S_a-v^{-1}\S_a\n^av\bigr),\eqno(32)$$
$$\d=0.\eqno(33)$$
Moreover,  a rather lengthy and tedious calculation also yields
$$\b=-v^{-2}\omega_a\n^a\P,\eqno(34)$$
$$\g=v^{-2}\omega_a\n^a\O.\eqno(35)$$
Hence, as a direct consequence of  eqs. (31) -
(35), we obtain
$$\n^a\bigl[2v^{-2}\xi_{(a}\S_{b)}\bigr]=v^{-2}\Bigl[\bigl(\n^a\S_a-
v^{-1}\S^a\n_av\bigr)\xi_b+(\omega_a\n^a\O)
(\n_b\P) + (\omega_a\n^a\O) (\n_b\O)\Bigr],\eqno(36)$$ which
implies along with eq. (27)
$$\eqalign{\n^aT_{ab}={v^{-1}\over
4\pi}\Bigl\{&\bigl[\n^a\n_a\P-v^{-1}\bigl\{(\n^a\P)(\n_av)
+ \o_a\n^a\O\bigr\} \bigr](\n_b\P)\cr+&
\bigl[\n^a\n_a\O-v^{-1}\bigl\{(\n^a\O)(\n_av)-\o_a\n^a\P
\bigr\}\bigr](\n_b\O)+
\n^a\bigl(v^{-1}\S^a\bigr)\xi_b\Bigr\} .\cr}\eqno(37)$$
 Since  the  vectors  $\xi^a,\n^a\P$  and  $\n^a\O$  are  linearly
independent  in  the  selected  region $\n^aT_{ab}=0$ implies that
eqs.  (22) and (23) must be satisfied.  This observation completes
the proof of our proposal.

 Finally, we would  like to emphasize  that whenever $\n^a\P$  and
$\n^a\O$  are  linearly  independent  the twice contracted Bianchi
identity gives an additional restriction on the possible behaviour
of  electromagnetic  fields.   Clearly,  the  above  argument also
yields that the coefficient of $\xi_b$ in eq.  (37) has to  vanish
identically, which means that the vector field,  $J^a=v^{-1}\S^a$,
is divergence  free.  Note,  that the  presence of  this conserved
current  is  consistent  with  the  general  result,  namely, if a
spacetime admits a  Killing vector field,  $\xi^a$, then eq.   (2)
can     be     integrated     to     the     conservation      law
$$\n^a(T_{ab}\xi^b)=0.\eqno(39)$$ For the particular matter  field
considered  in  this  paper  the  conserved current, $J^a$, can be
shown  to  be  proportional,  for  the  case of a timelike Killing
field, to the Poynting vector measured by observers following  the
Killing orbits.  When, in turn, $\xi^a$ is, for example, an  axial
Killing field the current is  proportional to one of the  flows of
the angular momentum of the electromagnetic field.  In both  cases
the divergence  free character  of $J^a$  leads to  the conclusion
that there  is no  net flux  of the  relevant current  through any
closed surface.

\bigskip
\bigskip
\parindent 0pt
{\bf References}
\bigskip
\item{[1]} A. Trautman: {\it Foundation and Current Problems of
General Relativity}, in {\sl Lectures on General Relativity},
(Brandeis Summer Institute in Theoretical Physics, Prentice -
Hall, Inc. Englewood Cliffs, New Yersey, 1965)

\item{[2]} R. M. Wald: {\it General Relativity}, (The University
of Chicago Press, Chicago, 1984)

\item{[3]} B. K. Harrison: J. Math. Phys. {\bf 9}, 1744 (1968)

\item{[4]} D. Kramer, H. Stefani, M. MacCallum and E. Herlt: {\it Exact
solutions of Einstein's equations} (Cambridge; Cambridge University Press,
1980)

\vfill\eject\end